\documentclass[
 reprint,
superscriptaddress,
 amsmath,amssymb,
 aps,
pra,
longbibliography
]{revtex4-1}

\usepackage{graphicx}
\usepackage{dcolumn}
\usepackage{bm}
\usepackage{bbm}
\usepackage[utf8]{inputenc}
\usepackage{amsfonts}
\usepackage{amsthm}
\usepackage{tikz}
\usepackage{bm}
\usepackage{physics}
\usepackage{comment}
\usepackage{xcolor}
\usepackage{blindtext}
\usepackage{pgfplots,pgfplotstable}
\usepgfplotslibrary{polar}
\usepgfplotslibrary{colormaps}
\pgfplotsset{compat=newest}
\usepackage{hyperref}

\definecolor{DTgreen}{rgb}{.1, .5, .2}

\definecolor{HS}{rgb}{0.0,0.5,0.0}
\definecolor{SN}{rgb}{1, 0.5, 0.31}
\definecolor{SSN}{rgb}{0.87, 1.0, 0.0}
\definecolor{OSN}{rgb}{1.0, 0.44, 0.37}
\newcommand{\re}[1]{\Re[#1]}
\newcommand{\im}[1]{\Im[#1]}

\renewcommand{\O}{O}
\newcommand{\I}{\mathcal{I}}
\newcommand{\vac}{\ket{\mathrm{vac}}}
\providecommand{\abs}[1]{\left\lvert#1\right\rvert}

\renewcommand{\d}{\mathrm{d}}

\newcommand{\e}{\mathrm{e}}
\newcommand{\ii}{\mathrm{i}}
\newcommand{\diag}{\mathrm{diag}}
\newcommand{\Vi}{\mathcal{F}_{\mathrm{in}}}
\newcommand{\Vo}{\mathcal{F}_{\mathrm{out}}}
\newcommand{\hVi}{\hat{\mathcal{F}}_{\mathrm{in}}}
\newcommand{\hVo}{\hat{\mathcal{F}}_{\mathrm{out}}}
\newcommand{\UBS}[1]{U_\mathrm{BS}(#1)}
\newcommand{\UPS}[1]{U_\mathrm{PS}(#1)}

\newcommand{\pBS}{\varphi_1}
\newcommand{\phicl}[1]{\varphi_{#1,\mathrm{cl}}}

\newcommand{\vphi}{\boldsymbol{\varphi}}
\newcommand{\vomega}{\boldsymbol{\omega}}
\newcommand{\vphicl}{\vphi_{\mathrm{cl}}}

\newcommand{\Uphi}{U_{\vphi}}
\newcommand{\hUphi}{\hat{U}_{\vphi}}

\newcommand{\uphi}{u_{\vphi}}

\newcommand{\Var}{\mathrm{Var}}

\theoremstyle{theorem}

\theoremstyle{definition}

\begin{document}

\title{Heisenberg scaling precision in the estimation of functions of parameters}

\author{Danilo Triggiani}
\email{danilo.triggiani@port.ac.uk}
\affiliation{School of Mathematics and Physics, University of Portsmouth, Portsmouth PO1 3QL, UK}

\author{Paolo Facchi}
\affiliation{Dipartimento di Fisica and MECENAS, Università di Bari, I-70126 Bari, Italy }
\affiliation{INFN, Sezione di Bari, I-70126 Bari, Italy}

\author{Vincenzo Tamma}
\email{vincenzo.tamma@port.ac.uk}
\affiliation{School of Mathematics and Physics, University of Portsmouth, Portsmouth PO1 3QL, UK}
\affiliation{Institute of Cosmology and Gravitation, University of Portsmouth, Portsmouth PO1 3FX, UK}

\date{\today}

\begin{abstract}
We propose a metrological strategy reaching Heisenberg scaling precision in the estimation of functions of any number $l$ of arbitrary parameters encoded in a generic $M$-channel linear network. This scheme is experimentally feasible since it only employs a single-mode squeezed vacuum and homodyne detection on a single output channel. Two auxiliary linear network are required and their role is twofold: to refocus the signal into a single channel after the interaction with the interferometer, and to fix the function of the parameters to be estimated according to the linear network analysed. Although the refocusing requires some knowledge on the parameters, we show that the required precision on the prior measurement is shot-noise, and thus achievable with a classic measurement. We conclude by discussing two paradigmatic schemes in which the choice of  the auxiliary stages allows to change the function of the unknown parameter to estimate.
\end{abstract}

\pacs{Valid PACS appear here}

\maketitle

\section{Introduction}

The estimation of physical properties has always played a central role in the development of science, engineering, technologies, and ultimately human knowledge. At the same time, advances in technologies and a better understanding of nature allow for improvements in the sensing protocols and, occasionally, for breakthroughs on the ultimate precisions fundamentally achievable. One of the most recent breakthroughs is the discovery of the advantage that quantum strategies can bring to metrology: it has been shown that the ultimate precision achievable employing $N$ entangled probes for the estimation of a single phase (or more in general the amplitude of a unitary evolution) exceeds the precision of any classical strategy~\cite{Giovannetti2004, Giovannetti2006, Dowling2008, Giovannetti2011}. In particular, it is possible to conceive quantum estimation strategies which lead to errors that scale as fast as $1/N$, bound usually called \textit{Heisenberg Limit} (HL), whereas the error of any classical strategy is bounded by the so-called \textit{Shot-Noise Limit} (SNL) and cannot scale faster than $1/\sqrt{N}$.
Since these initial works, considerable effort has been put in the development of quantum estimation protocols reaching the HL~\cite{Dowling2015, DePasquale2015, Zhou2018, Ge2018,Qian2019}, with applications in imaging~\cite{McConnell2017, Unternahrer2018}, thermometry~\cite{DePasquale2018,Seah2019}, magnetic field~\cite{Razzoli2019, Bhattacharjee2020} and gravitational waves detection~\cite{Ligo2013}, among others. 

A difficulty often encountered in early protocols is the fragility of the required probes state, which are usually entangled. 
In order to overcome this drawback, the employment of Gaussian states as probes and \textit{squeezing} as resource has recently started to be considered, since these states are feasible to be experimentally produced and robust against noise~\cite{Monras2006, Pezze2008, Aspachs2009, Lang2013, Maccone2019, Matsubara2019, Oh2019, Gatto2019, Gatto2020, Gramegna2020Typicality, Gramegna2020Letter}. 
The development of the so-called \textit{Gaussian metrology}, and thus the possibility to conceive easy-to-implement estimation protocols that are still able to reach the full quantum advantage, have recently stimulated the study of quantum strategies for the more complex case of estimation of functions of parameters, with applications in field-gradients sensing or non-linear functions interpolation~\cite{Zhuang2018, Ge2018, Qian2019, Gatto2019, Gatto2020, Xia2020, Guo2020}. 
However, despite their feasibility, previous protocols still require a number of constraints on the parameters and the setups, such as commuting generators, reduced working range, scarce freedom on the structure of the networks and on the nature of the parameters, which need to be overcome for practical applications, including scenarios where small or no control is possible on the structure and parameter dependence of a given arbitrary multi-parameter network.

\begin{figure}[t]
\centering
\pgfplotsset{
colormap={whitered}{color(0cm)=(blue!60); color(1cm)=(red!90)}
}

\includegraphics[width=.45\textwidth]{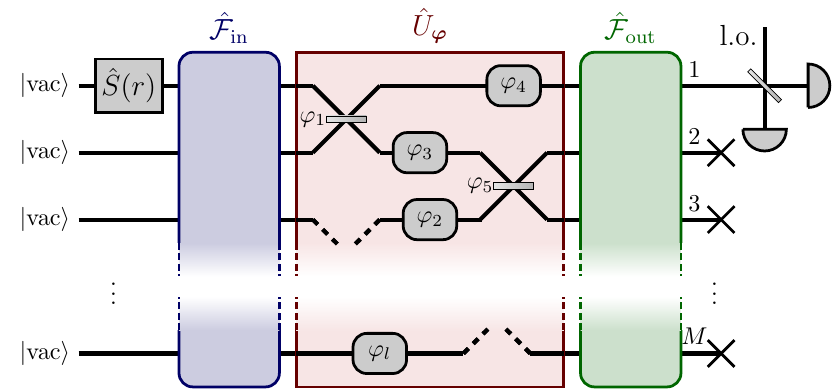}
	\caption{Optical scheme for Heisenberg limited estimation of a function $f(\vphi)$ of an arbitrary number of parameters $\varphi_i$, with $i = 1\dots l$, encoded  in  an arbitrary $M$-channel linear optical network $\hUphi$ by using either two auxiliary optical stages $\hVi$ and $\hVo$ or only one of them. Such parameters can be optical phases or phase-like parameters of beam-splitters. By injecting a single-mode squeezed-vacuum state with $N$ mean photons in (say) the first channel, through single-mode homodyne detection in (say) the first channel and ignoring the others, it is possible to tune the phase of the local oscillator (l.o) to estimate the phase acquired by the probe through the whole network, which is a function of the $l$ parameters and of the specifics of the auxiliary stages $\hVi$ and $\hVo$. When an auxiliary stage ($\hVi$ or $\hVo$) is suitably chosen, this setup allows Heisenberg scaling precision in the photon number $N$. Furthermore, the extra degrees of freedom in either of the two auxiliary gates could be used to manipulate the functional dependence of $f(\vphi)$ on the parameters.}
	\label{fig:Generic}
\end{figure}

Differently from these previous approaches, we demonstrate Heisenberg limited estimation of an arbitrary number of either linear or non-linear functions of $l>1$  independent parameters, which can be encoded in an arbitrary manner in any passive linear  optical network. 
Indeed, for any given parameter-dependent network, the function of the parameters to be estimated can be manipulated by using simple linear optical stages (see \figurename~\ref{fig:Generic}). 
Furthermore, no constraint in the values of the parameters and no entangled probes are needed. 
In particular, the $l$ unknown parameters can be  both optical phases or reflectivities of beam-splitters which can depend in principle also from external parameters such as pressure, electromagnetic fields or temperature.
The input probe is a single-mode squeezed vacuum, and homodyne measurement on a single output channel is performed, while two auxiliary linear network are employed before and after the interferometer to scatter first and then refocus the photons.
 Although these auxiliary stages might depend on the value of the unknown parameters, requiring thus a prior knowledge in order to correctly implement our scheme, we show that shot-noise precision on the prior knowledge of the parameters is sufficient, meaning that a prior classical measure of each parameter is enough to correctly select the auxiliary stages. 
Moreover, for a given interferometer, these stages are not unique, and it is possible to employ this freedom to manipulate the function of the parameters that is possible to estimate at Heisenberg scaling precision. 
Finally, we apply our general scheme in the case of two different multi-parameter interferometers: the first is the simplest non-trivial example of multi-parameter estimation, which allows to estimate a whole family of (in general) nonlinear functions:  a linear interferometer with three parameters encoded in two phase-shifts and a beam-splitter; the second allows to estimate any linear combination with positive coefficients of phase-shifts and beam-splitter parameters.

\section{Linear network with an arbitrary number of unknown parameters} \label{sec:Setup}

Let us consider an $M\times M$ passive and linear optical network which depends on $l$ unknown parameters that, in a compact form, can be written as component of an $l$-dimensional vector $\vphi = (\varphi_1,\dots,\varphi_l)$. The network is described by a unitary operator $\hUphi$ which depends smoothly on $\vphi$, whose action on the $M$ bosonic annihilation operators $\hat{a}_i$, for $i=1,\dots,M$ is given by the unitary matrix $\Uphi$ such that
\begin{equation}
\hUphi^\dag \hat{a}_i \hUphi = \sum_{j=1}^M (\Uphi)_{ij}\hat{a}_j.
\end{equation}

The input probe in our scheme consists in a single-mode squeezed-vacuum state with $N = \sinh^2 r$ mean number of photons, where $r$ is the real squeezing parameter of the probe. This is injected in one of the input ports, namely the first, of an auxiliary linear and passive optical network $\hVi$ (described by the unitary matrix $\Vi$), whose role is to distribute the probe photons among all the channels of the network $\hUphi$, and ultimately to control the function of the parameters $f(\vphi)$ to estimate. 
Then, the photons go through a second auxiliary linear and passive network $\hVo$, described by $\Vo$, which refocuses the photons into a single output port, say the first, so that all the information on the parameters acquired by the probe can be read by performing homodyne detection on a single channel. The probability amplitude  $\chi_{\vphi}$ associated with the transition of a single photon from the first input port to the first output port of the whole setup reads
\begin{equation}
\chi_{\vphi} \equiv \sqrt{P_{\vphi}}\e^{\ii f(\vphi)} = (\Vo \Uphi \Vi)_{11}
\label{eq:GenericProbAmp}
\end{equation}
and represents the only relevant quantity in the presented scheme. 
If the refocusing process is perfect, the probability $P_{\vphi}$ will be exactly equal to one, and all the  information on $\vphi$ is encoded in the phase acquired by the probe $f(\vphi)$, whose functional dependence on the parameters $\vphi$ can be partially controlled through the auxiliary gates $\hVi$ and $\hVo$. 
In the next section we will show that Heisenberg limited sensitivity can be achieved even for an imperfect refocusing as long as the probability for the photons to exit from a different channel scales as $1/N$, which is attainable without any additional quantum resource.

Since the initial squeezing parameter is real, and the probe acquires a total phase-shift of $f(\vphi)$, the quadrature field with minimum variance at the first output port of the interferometer corresponds to $\hat{x}_{f(\vphi)\pm\pi/2}$. Eventually, the quadrature field $\hat{x}_\theta$ is measured through homodyne detection on the first channel, where $\theta$ is the reference phase of the homodyne local oscillator, in order to infer the value of $f(\vphi)$. 
From equation~\eqref{eq:GenericProbAmp} it is possible to make explicit the dependence of the acquired phase $f(\vphi)$ from the elements of the scattering matrix $\Uphi$ 
\begin{align}
f(\vphi) = \arctan(\frac{\im{(\Vo \Uphi \Vi)_{11}}}{\re{(\Vo \Uphi \Vi)_{11}}})
\end{align}
and the influence of both the networks $\Vi$ and $\Vo$. In the next section, we will show that the precision reached with this setup in the estimation of the function $f(\vphi)$ of the parameters asymptotically reaches the Heisenberg limit.

\section{Heisenberg limited estimation of a function of the network parameters}

The probability density function $p(x|\vphi)$ that the outcome $x_\theta$ of the homodyne detection falls between $x$ and $x + \dd x$
is a centred Gaussian distribution
\begin{equation}
p(x|\vphi)= \frac{1}{\sqrt{2\pi\sigma^2_{\vphi}}}\exp[-\frac{x^2}{2\sigma^2_{\vphi}}],
\label{eq:GaussianProbability}
\end{equation}
whose variance
\begin{align}
\sigma^2_{\vphi} = \frac{1}{2} + P_{\vphi}\left(\sinh^2 r + \cos(2f(\vphi)-2\theta)\sinh r\cosh r\right)
\label{eq:VarianceGeneric}
\end{align}
depends on the parameters $\vphi$ through $P_{\vphi}$ and $f(\vphi)$ (see appendix~\ref{app:ProbabilityDistribution}). Due to the presence of multiple unknown parameters in the optical network $U_{\vphi}$, the ultimate precision achievable on the estimation of any function $\alpha(\vphi)$ of the parameters is regulated by the Fisher information matrix~\cite{cramer1999mathematical}
\begin{equation}
\I = \int \dd x\, p(x|\vphi)\, \bigl(\vnabla_{\vphi} \log p(x|\vphi)\bigr) \, \bigl(\vnabla_{\vphi} \log p(x|\vphi) \bigr)^\mathrm{T}
\end{equation}
where $\vnabla_{\vphi} = (\partial_{\varphi_1},\dots,\partial_{\varphi_l})^{\mathrm{T}}$ is the gradient in parameter space. In particular, substituting the distribution~\eqref{eq:GaussianProbability} in this expression, we obtain
\begin{equation}
\I = \frac{1}{2\sigma^4_{\vphi}}\left(\vnabla_{\vphi}\sigma^2_{\vphi}\right)\left(\vnabla_{\vphi}\sigma^2_{\vphi}\right)^{\mathrm{T}}.
\label{eq:FisherMatrixGauss}
\end{equation}

We assume now that the refocusing performed by the stage $\hVo$ is such that the probability $P_{\vphi}$ differs from  unity by a small quantity of order $\O(N^{-1})$, namely
\begin{equation}
P_{\vphi} \sim 1 - \frac{\ell}{N},\quad \ell\geq0
\label{eq:Condition1}
\end{equation}
with $l$ independent of $N$. For an arbitrary $\vphi$ dependent network $\hUphi$ this condition can only be satisfied with a prior knowledge of the parameters in order to suitably chose the auxiliary gates. Remarkably, in appendix \ref{app:Prior} we show that a classical prior knowledge $\vphicl$ of the parameters, corresponding to a shot-nose precision $\delta \vphi =\vphi-\vphicl = \O(N^{-1/2})$, suffices to satisfy \eqref{eq:Condition1}.  

Moreover, we impose that the local oscillator phase $\theta$ is experimentally tuned on a value $\theta_{\vphi}$ of the asymptotic form
\begin{equation}
\theta_{\vphi} \sim f(\vphi) \pm \frac{\pi}{2} + \frac{k}{N},\quad k\neq 0
\label{eq:Condition2}
\end{equation}
which differs from the phases $f(\vphi) \pm \pi/2$ of the quadrature field with minimum variance, only by a quantity $k/N$, with $k$ independent of $N$. When conditions \eqref{eq:Condition1} and \eqref{eq:Condition2} hold, in the large $N$ limit, the Fisher information matrix reads (see appendix~\ref{app:FisherMatrix})
\begin{equation}
\I \sim 8\varrho(k,\ell)N^2 \bigl( \vnabla_{\vphi}f(\vphi) \bigr) \bigl(\vnabla_{\vphi}f(\vphi) \bigr)^\mathrm{T},
\label{eq:FisherMatrix}
\end{equation}
with 
\begin{equation}
\varrho(k,\ell) = \left(\frac{8k}{1+16k^2+4\ell}\right)^2.
\end{equation}
A thorough analysis of the matrix $\I$ in~\eqref{eq:FisherMatrix} (see appendix~\ref{app:CRB}) shows  that only $f(\vphi)$, or functions of $f(\vphi)$, admit estimators with finite variances.
In particular, any unbiased estimator $\tilde{f}$ of $f(\vphi)$ is characterized by a variance which satisfies (see appendix~\ref{app:CRB})
\begin{equation}
\Var(\tilde{f}) \geq \frac{1}{8\varrho(k,\ell)N^2},
\end{equation}
and thus our scheme allows Heisenberg scaling precision in $N$.

From equation~\eqref{eq:GenericProbAmp} we notice that the functional dependence of $f(\vphi)$ on the parameters $\vphi$ changes for different choices of $\hVi$ and $\hVo$. In particular, after the optimization required to satisfy condition~\eqref{eq:Condition1}, the remaining degrees of freedom on the stages $\hVi$ and $\hVo$ can be employed to manipulate the function of the parameters $f(\vphi)$ that we can estimate with Heisenberg scaling precision. Hereafter we are going to show some examples for which the acquired phase $f(\vphi)$ to be estimated assumes simple functional dependences of the parameters $\vphi$, which can be manipulated through simple changes in the two auxiliary stages, useful for practical purposes.

\section{Examples of setups for the estimation of functions of parameters}
\subsection{Functions of parameters in a two-channel network}
\label{sec:Example2Channels}

\begin{figure}[h]
	\centering
	
\includegraphics[width=.45\textwidth]{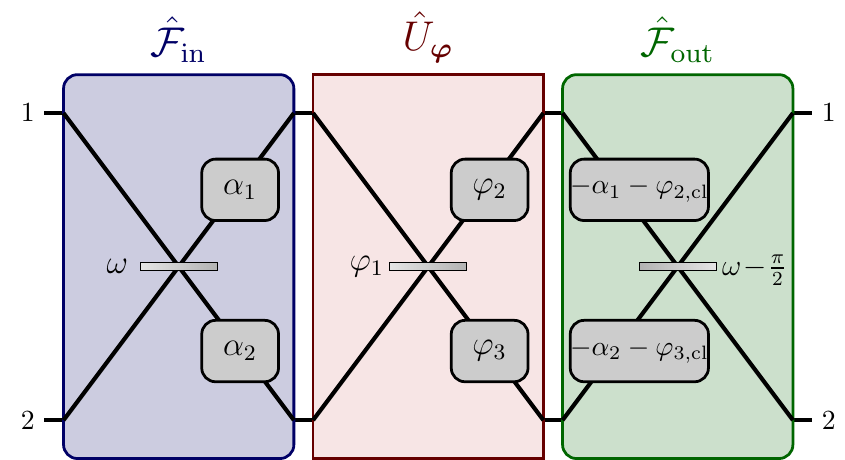}
	\caption{Example of an $M$-channel linear network $\hUphi$ in \ref{fig:Generic} for $M=2$ with suitable auxiliary stages $\hVi$ and $\hVo$ for the the estimation of a family of functions of two optical phases and a beam-splitter parameter within $\hUphi$. The function of the parameters $\vphi\equiv(\varphi_1,\varphi_2,\varphi_3)$ estimated depends on a control parameter $\Delta\alpha = \alpha_1-\alpha_2$ which can be arbitrarily chosen. In order to correctly tune the auxiliary stage $\hVo$ to achieve Heisenberg limited sensitivity, only a classical prior knowledge $\vphicl$ on the parameter is required, namely the error $\delta\vphi = \vphi - \vphicl$ in a preparative coarse estimation must be of order of $1/\sqrt{N}$.}
	\label{fig:Example1}
\end{figure}

We consider a $2$-modes network (see Fig.~\ref{fig:Example1}), which allows a Heisenberg scaling precision in the estimation of a family of functions $f(\vphi;\Delta\alpha)$ of the reflectivity $\pBS$ of a beam splitter, described by a scattering matrix 
\begin{equation}
\UBS{\pBS}=\e^{\ii\varphi_1\sigma_y},
\end{equation}
with $\sigma_y$ being the second Pauli matrix, and of two phase-shifts described by \begin{equation}
\UPS{\varphi_2,\varphi_3}= 
\begin{pmatrix}
\e^{\ii \varphi_2} & 0 \\
0 &  \e^{\ii \varphi_3}
\end{pmatrix}.
\end{equation} The family of functions is parametrized by the relative phase shift $\Delta\alpha=\alpha_1 - \alpha_2$ in the arms of an input auxiliary gate $\hVi\equiv\hVi(\alpha_1,\alpha_2)$, where $\alpha_1$ and $\alpha_2$ can be arbitrarily chosen.
The probe photons, in a single-mode squeezed state $\hat{S}_1(r)\vac = \e^{\frac{r}{2}(\hat{a}_1^2 - \hat{a}_1^{\dag 2})}\vac$ with an average number of photons $N = \sinh^2 r$, are firstly injected into the first channel of the auxiliary input linear network $\hVi(\alpha_1,\alpha_2)$ which, in general, may require some prior classical knowledge $\phicl{1}$ of the  beam-splitter parameter $\pBS$, namely such that the error $\delta\pBS = \pBS - \phicl{1}$ in the prior coarse estimation is of order $\O(1/\sqrt{N})$. In particular, in order to satisfy condition~\eqref{eq:Condition1}, a possible choice for this stage consists in a beam splitter $\UBS{\omega}$, whose reflectivity  $\omega$ is tuned according to 
\begin{equation}
\omega = \frac{1}{2} \arctan \left(\frac{\cos(\phicl{1})}{\sin(\phicl{1})\cos\Delta\alpha}\right),
\label{eq:OmegaDefinition}
\end{equation}
and in the two arbitrary and $\vphi$-independent phase shifts $\alpha_1$ and $\alpha_2$, so that the scattering matrix of this stage is \begin{equation}
\Vi=\UPS{\alpha_1, \alpha_2}\,\UBS{\omega}.
\end{equation} 
Then, the probe goes through the passive linear network described by the matrix 
\begin{equation}
\Uphi=\UBS{\varphi_1} \, \UPS{\varphi_2,\varphi_3},
\end{equation} 
and finally through a second auxiliary linear network $\hVo(\alpha_1,\alpha_2)$ whose preparation requires a prior classical knowledge $\vphicl$ on the three unknown parameters, namely such that the errors $\delta\varphi_i=\varphi_i - \phicl{i}$ for $i=1,2,3$ in the prior estimations are of order of $\O(1/\sqrt{N})$. In particular, this stage is composed of a phase-shift in each channel of values $-\alpha_1-\phicl{2}$ and $-\alpha_2-\phicl{3}$, and of a beam-splitter with reflectivity $\omega-\pi/2$. The scattering matrix of the whole output stage thus reads \begin{equation}
\Vo(\alpha_1,\alpha_2) = \UBS{\omega-\pi/2}\UPS{-\alpha_1-\phicl{2},-\alpha_2-\phicl{3}} .
\end{equation} 
Finally, a homodyne detection is performed on the first output port of the interferometer according to condition~\eqref{eq:Condition2}.

A straightforward calculation shows that, for this setup, the one-photon transition amplitude~\eqref{eq:GenericProbAmp} reads
\begin{align}
\chi_{\vphi} &= \e^{\ii\frac{\delta\varphi_2 + \delta\varphi_3}{2}}\notag\\
&\times\Bigg(\cos(\frac{\delta\varphi_2-\delta\varphi_3}{2})\sin 2\omega\cos\varphi_1\notag\\
&\qquad +\cos(\Delta\alpha-\frac{\delta\varphi_2-\delta\varphi_3}{2})\cos 2\omega \sin\varphi_1 \notag\\
&\qquad + \ii \sin(\Delta\alpha-\frac{\delta\varphi_2-\delta\varphi_3}{2})\sin\varphi_1\Bigg),
\label{eq:ProbAmp2}
\end{align}
and is such that condition~\eqref{eq:Condition1} on the transition probability $P_{\vphi}=\abs{\chi_{\vphi}}^2$ is satisfied for a reflectivity $\omega$ given by~\eqref{eq:OmegaDefinition}. The acquired phase $f(\vphi;\Delta\alpha)$ through the interferometer reads
\begin{widetext}
\begin{equation}
f(\vphi;\Delta\alpha) = \frac{\delta\varphi_2+\delta\varphi_3}{2} + \arctan(\frac{\sin\varphi_1\sin(\Delta\alpha-\frac{\delta\varphi_2-\delta\varphi_3}{2})\sqrt{1-\sin^2(\phicl{1})\sin^2\Delta\alpha}}{\cos\varphi_1\cos\phicl{1}\cos(\frac{\delta\varphi_2-\delta\varphi_3}{2})+\sin\varphi_1 \sin\phicl{1}\cos\Delta\alpha \cos(\Delta\alpha - \frac{\delta\varphi_2-\delta\varphi_3}{2})}),
\label{eq:f2modes}
\end{equation}
\end{widetext}
and in general is not linear in the parameters $\vphi$. It can be arbitrarily tuned by changing the value of $\Delta\alpha$ and estimated with Heisenberg scaling precision through homodyne measurement and Bayesian analysis~\cite{olivares2009,berni2015,Pezze2008}.

For certain choices of $\Delta\alpha$, the  function $f(\vphi;\Delta\alpha)$ becomes linear in $\vphi$. For example, for $\Delta\alpha=\pi/2$, the beam splitters needed in the input stage $\hVi$ and output stage $\hVo$ are balanced, since  from condition~\eqref{eq:OmegaDefinition} one gets $\omega= \pm \pi/4$. For $\omega=+\pi/4$, the acquired phase becomes 
\begin{equation}
 f(\vphi;\pi/2)=\varphi_1 + \frac{\delta\varphi_2+\delta\varphi_3}{2}.
 \label{eq:linearf}
\end{equation}
In particular, if both $\varphi_2$ and $\varphi_3$ are perfectly known (i.e. $\varphi_2 = \phicl{2}$ and $\varphi_3 = \phicl{3}$) so that $\delta\varphi_2=\delta\varphi_3=0$, then the network in Fig.~\ref{fig:Example1} reduces to the one in Fig.~\ref{fig:Example2}(b), and  can be employed for the estimation of $\varphi_1$, namely the parameter associated to the beam-splitter, without requiring any prior information.
 
For $\Delta\alpha=0$ instead, condition~\eqref{eq:OmegaDefinition} reads $\omega=\pi/4 - \phicl{1}/2$, and the phase acquired becomes 
\begin{align}
f(\vphi;0)&=\frac{\delta\varphi_2+\delta\varphi_3}{2} 
\nonumber\\
& \quad + \arctan(\tan(\frac{\delta\varphi_2-\delta\varphi_3}{2})\frac{\sin(\varphi_1)}{\cos(\delta\varphi_1)})\notag\\
&=\frac{\delta\varphi_2+\delta\varphi_3}{2} + \frac{\delta\varphi_2-\delta\varphi_3}{2}\sin(\varphi_1) + \O(N^{-3/2}),
\label{eq:sinf}
\end{align}
where we exploited the fact that the errors $\delta\varphi_2$ and $\delta\varphi_3$ are of order $\O(1/\sqrt{N})$. Noticeably, this is an example of estimation of a non-linear function of the unknown parameters. One might think to employ this setup when the phase-shifts $\varphi_2$ and $\varphi_3$ are known, and purposely tune off $\phicl{2}$ and $\phicl{3}$ in $\hVo$ of certain and known quantities $\delta\varphi_2\neq\delta\varphi_3$, e.g. $\delta\varphi_2=-\delta\varphi_3=\lambda$. In this case, provided that $\lambda$ is kept of order $\O(1/\sqrt{N})$, this network can be optimally employed to estimate directly with Heisenberg scaling precision $f(\vphi;0)=\lambda\sin\varphi_1$, namely the transmittivity amplitude of the unknown beam splitter multiplied by a known factor $\lambda$. 

If instead $\varphi_1=0$, this setup reduces to a Mach-Zehnder interferometer with balanced beam-splitters, and the function~\eqref{eq:f2modes} becomes the average of the two remaining unknown parameters $\delta\varphi_2$ and $\delta\varphi_3$. In the following section, we will show a generalization of this case, in which an $M$-channel Mach-Zehnder-like interferometer can be exploited for the estimation of an arbitrary linear combination with positive weights of phase-shifts, or beam-splitter reflectivities as displayed in~\eqref{eq:linearf} when $\varphi_2$ and $\varphi_3$ are known.

\subsection{Linear combination of any number of phase-shifts and of beam-splitter reflectivities }
\begin{figure}[h]
	\centering
	
\includegraphics[width=.45\textwidth]{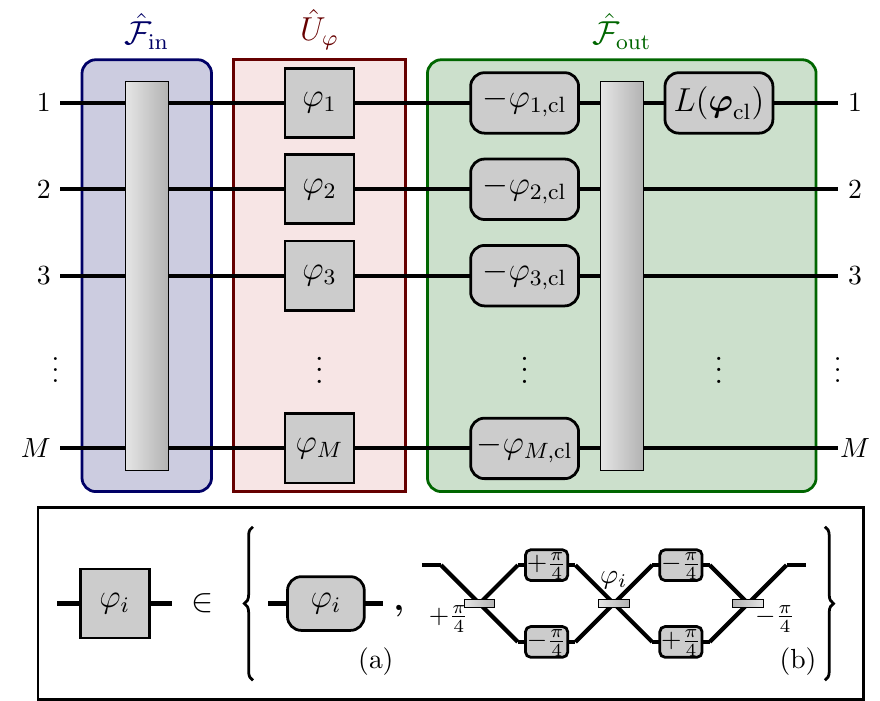}
	\caption{An $M$-channel linear and passive network which allows Heisenberg scaling precision in the estimation of any linear combination of $M$ unknown parameters $\vphi$, all encoded within $\Uphi$ (red box). As shown in the lower panel, each parameter $\varphi_i$ is (a) an optical phase acquired through a single-mode phase-shift, or (b) the reflectivity of a lossless beam-splitter which can be readily imprinted into an optical phase by means of a $2$-channel parameter-independent local network. Since the probability that a photon injected in the first port of each local network comes out from the second channel is exactly zero, each local network acts as a single-channel phase delay with magnitude $\varphi_i$.}
	\label{fig:Example2}
\end{figure}

We finally show an interferometer (see Fig.~\ref{fig:Example2}) which allows Heisenberg scaling precision in the estimation of any linear combination
\begin{equation}
L(\vphi)=\sum_{i=1}^M \omega_i \varphi_i
\label{eq:LinComb1}
\end{equation}
of $M$ independent and unknown quantities $\vphi=(\varphi_1,\dots,\varphi_M)$, which can be both optical phases acquired through single-channel phase-shifts $\UPS{\varphi_i}=\e^{\ii\varphi_i}$, or reflectivities of beam splitters $\UBS{\varphi_i}=\e^{\ii\varphi_i\sigma_y}$, with $M$ non-negative coefficients ${\vomega=(\omega_1,\dots,\omega_M)}$. In our discussion, we will additionally suppose that the positive coefficients $\vomega$ sum to one, without any loss of generality. 

As previously discussed, the probe employed is a single-mode squeezed state $\hat{S}_1(r)\vac$, with $N=\sinh^2 r$ average number of photons. The probe is injected in the first port of a first auxiliary  linear network $\hVi$ which scatters the photons into each channel of $\hUphi$ with probabilities
\begin{equation}
\abs{(\Vi)_{i1}}^2 = \omega_i,
\end{equation}
where the normalization of the positive coefficients allows for the unitarity of $\Vi$. The network $\hUphi$ encodes all the unknown parameters $\vphi$, each one associated with a different channel. For each beam-splitter with an unknown reflectivity, an auxiliary $2$-channel $\vphi$-independent network $V$ (See Fig.~\ref{fig:Example2}(b)) is employed in order to turn its parameter into an optical phase. In particular, the network 
\begin{equation}
V^\dag\,\UBS{\varphi}\,V=\UPS{\varphi,-\varphi},
\label{eq:localN}
\end{equation} 
with 
\begin{equation}
V=\UPS{\pi/4,-\pi/4}\, \UBS{\pi/4},
\end{equation} 
acts as a phase shift on both channels. Thus, when a signal is fed into the first port of this network, all the light comes out from the first output port shifted by a phase $\varphi$, so it behaves as a single-channel phase-shift of magnitude $\varphi$. Noticeably, the network $V$ needed for this purpose is the same employed in Sec.~\ref{sec:Example2Channels}, in the case of $\varphi_2=\varphi_3=0$ for $\Delta\alpha=\pi/2$. It is worth mentioning that if the signal is injected in the second channel of this local network, it will come out only from the second output channel shifted by $-\varphi$, hence allowing in this way a negative weight in the linear combination~\eqref{eq:LinComb1}. At the output of $\hUphi$, a second auxiliary passive and linear network $\hVo$ is employed, whose preparation requires a prior classical knowledge $\vphicl=(\phicl{1},\dots,\phicl{M})$ of the unknown parameters, meaning that the errors $\delta\varphi_i=\varphi_i - \phicl{i}$ are of order of $\O(1/\sqrt{N})$. In particular, the signal in the $i$-th channel must undergo a phase shift of $-\phicl{i}$. Then, the probe is refocused on a single channel by inverting the action of $\hVi$, and it undergoes a phase-shift of constant and known magnitude $L(\vphicl)$, so that the scattering matrix of the output stage reads 
\begin{equation}
\Vo = \UPS{L(\vphicl),0,\dots,0}\, \Vi^\dag\, \UPS{-\vphicl},
\end{equation} 
where we denoted with $\UPS{\lambda_1,\dots,\lambda_l}=\diag(\e^{\ii\lambda_1},\dots,\e^{\ii\lambda_l})$. Finally, single-mode homodyne detection is performed according to condition~\eqref{eq:Condition2}.

With this setup, the probability amplitude, shown in~\eqref{eq:GenericProbAmp}, reads
\begin{align}
\chi_{\vphi}&=\e^{\ii L(\vphicl)}\sum_{i=1}^M \omega_i\e^{\ii \delta\varphi_i}=\notag\\
&=\e^{\ii L(\vphicl)}\left(1+\sum_{i=1}^M \ii \omega_i\delta\phi_i - \frac{1}{2} \sum_{i=1}^M \omega_i\delta\phi_i^2\right) + \O(N^{-3/2}).
\label{eq:ProbAmpPhi}
\end{align}
The shot-noise scaling $\delta\phi_i = \O(N^{-1/2})$, implies that the one-photon probability
\begin{align}
P_{\vphi}&=\abs{\e^{\ii L(\vphicl)}\sum_{i=1}^M \omega_i \e^{\ii \delta\varphi_i}}^2=\notag\\
&= 1 + \left(\sum_{i=1}^M \omega_i\delta\varphi_i\right)^2 - \sum_{i=1}^M \omega_i\delta\varphi_i^2 + \O(N^{-3/2})\notag\\
&\equiv 1 - \frac{\ell}{N} + \O(N^{-3/2})
\end{align}
satisfies condition~\eqref{eq:Condition1}, so that Heisenberg scaling sensitivity in the estimation of the total acquired phase $f(\vphi)=\arg(\chi_{\vphi})$ can be achieved. In particular
\begin{align}
f(\vphi) &= L(\vphicl)+\sum\limits_{i=1}^M \omega_i\delta\varphi_i + \O(N^{-3/2})=\notag\\
&= L(\vphi) + \O(N^{-3/2}),
\label{eq:gammaExample1}
\end{align}
so that it is possible to recover the linear combination~\eqref{eq:LinComb1} with Heisenberg scaling precision from the estimation of $f(\vphi)$. Notice that, although $L(\vphi)$ in~\eqref{eq:LinComb1} and $f(\vphi)$ in~\eqref{eq:gammaExample1} are not exactly equal, they differ by a quantity of order $\O(N^{-3/2})$ which is beyond the Heisenberg resolution, and thus negligible for our estimation purposes.

We conclude this section with an insightful observation. As already discussed before, the local networks~\eqref{eq:localN} shown in Fig.~\ref{fig:Example2}(b), appearing inside $\Uphi$, whose purpose is to translate the unknown beam-splitter reflectivities into optical phase-shifts, can be obtained from the setup in Fig.~\ref{fig:Example1} by imposing $\Delta\alpha=\pi/2$, when $\varphi_2=\varphi_3=0$. It is indeed possible to further generalise this scheme by replacing the local networks~\eqref{eq:localN} with the more general network in Fig.~\ref{fig:Example1}, and only requiring conditions~\eqref{eq:OmegaDefinition} to hold for each local network (See appendix~\ref{app:Generalisation}). This allows to estimate with Heisenberg scaling precision linear combinations of functions of unknown local parameters of the type shown in~\eqref{eq:f2modes}, with each function parametrised by an arbitrary local quantity $\Delta\alpha_i$.

\section{Conclusions}

We provided a feasible metrologic strategy to estimate functions of multiple parameters encoded arbitrarily in a $M$-channel linear network with Heisenberg scaling precision in the average number of photons injected. 
Our scheme is experimentally feasible since it only requires a single-mode squeezed-vacuum state which, in general, is scattered by a first auxiliary network, interacts with the interferometer, and eventually is refocused by a second auxiliary network in a single output channel, where homodyne detection takes place. 
We showed that in order for the refocusing to be successful, a prior knowledge on the unknown parameters is required for the realization of only one of the two stages (i.e. $\hVo$) and only with a precision at shot-noise level, so that a classical estimation strategy is sufficient. 
Remarkably, the remaining degrees of freedoms in the stage $\hVo$ and all the ones in the parameter-independent stage, in this case $\hVi$, can be used to manipulate the functional expression of the unknown parameters depending on which function we are interested to estimate with Heisenberg limited precision. 
We provide as examples two interferometric schemes, and we demonstrate how the choice of the auxiliary stages influences the functions of the parameters that is possible to estimate with Heisenberg scaling precision: 
in the first, we examine a non-trivial two-channel linear network with unknown phase-shifts and beam-splitters, and we show how a whole family of functions, in general non-linear in the parameters and parametrized by an arbitrary relative phase in the auxiliary stages, can be estimated with Heisenberg scaling precision; 
in the second, we show that it is possible to estimate any linear combination of phase-shifts and beam-splitter reflectivities, with the only requirement for the linear combination coefficients to be positive. 
Our results are strongly relevant in experimental scenarios where we are interested to measure the global properties of a given network associated to particular functional dependencies from its parameters with no constraints on their values.

\section{Acknowledgements}
We thank Frank A. Narducci for useful discussions. This work was supported by the Office of Naval Research Global (N62909-18-1-2153). PF is partially supported by Istituto Nazionale di Fisica Nucleare (INFN) through the project “QUANTUM”, and by the Italian National Group of Mathematical Physics (GNFM-INdAM).

\appendix

\section{Variance of the quadrature $\hat{x}_\theta$.}\label{app:ProbabilityDistribution}

In order to evaluate the variance~\eqref{eq:VarianceGeneric} along the quadrature $\hat{x}_\theta$ of the squeezed vacuum state after the action of the interferometer, we firstly recall the covariance matrix $\Gamma_0$ of the input state $\hat{S}_1(r )\ket{\mathrm{vac}}$, which reads
\begin{equation}
\Gamma_0 =\dfrac{1}{2}
\begin{pmatrix}
\e^{2\mathcal{R}} & 0\\
0 & \e^{-2\mathcal{R}}
\end{pmatrix},
\end{equation}
where $\mathcal{R}$  is the $M\times M$ diagonal matrix $\mathcal{R} = \diag(r,0,\dots,0)$. After the action of the interferometer, the covariance matrix transforms into
\begin{equation}
\label{eq:AppGammaPhi}
\Gamma_{\vphi} = R_{\vphi} \Gamma_0 R_{\vphi}^T,
\end{equation}
where $R_{\vphi}$ is the orthogonal and symplectic matrix associated with the interferometer unitary matrix $\uphi=\Vo\Uphi\Vi$
\begin{equation}
R_{\vphi} =
\begin{pmatrix}
\re{\uphi} & -\im{\uphi}\\
\im{\uphi} & \re{\uphi}
\end{pmatrix}.
\end{equation}
Thus $\Gamma_{\vphi}$ in~\eqref{eq:AppGammaPhi} reads
\begin{equation}
\Gamma_{\vphi} =
\begin{pmatrix}
\Delta X^2_{\vphi} & \Delta XP_{\vphi}\\
\Delta XP_{\vphi}^T & \Delta P^2_{\vphi}
\end{pmatrix}.
\end{equation}
where 
\begingroup
\allowdisplaybreaks
\begin{align}
\Delta X^2_{\vphi} &\equiv  \dfrac{1}{2}\left[\re{\uphi}\e^{2\mathcal{R}}\re{\uphi^\dag} - \im{\uphi}\e^{-2\mathcal{R}}\im{\uphi^\dag}\right]\notag\\
&=\dfrac{1}{2}\left[\Re[\uphi\cosh(2\mathcal{R})\uphi^\dag]+\Re[\uphi\sinh(2\mathcal{R})\uphi^\mathrm{T}]\right],\\
\Delta P^2_{\vphi} &\equiv \dfrac{1}{2}\left[ -\im{\uphi}\e^{2\mathcal{R}}\im{\uphi^\dag} + \re{\uphi}\e^{-2\mathcal{R}}\re{\uphi^\dag}\right]\notag\\
&= \dfrac{1}{2}\left[\Re[\uphi\cosh(2\mathcal{R})\uphi^\dag]-\Re[\uphi\sinh(2\mathcal{R})\uphi^\mathrm{T}]\right],\\
\Delta XP_{\vphi} &\equiv \dfrac{1}{2}\left[ -\re{\uphi}\e^{2\mathcal{R}}\im{\uphi^\dag} - \im{\uphi}\e^{-2\mathcal{R}}\re{\uphi^\dag}\right]\notag\\
&=\dfrac{1}{2}\left[-\Im[\uphi\cosh(2\mathcal{R})\uphi^\dag]+\Im[\uphi\sinh(2\mathcal{R})\uphi^\mathrm{T}]\right].
\end{align}
\endgroup
In the second lines of each of the previous expression, we have used the fact that $\mathcal{R}$ is real. 

The $2\times 2$ reduced covariance matrix $\Gamma'_{\vphi}$ of the first mode reads
\begin{equation}
\Gamma'_{\vphi} =
\begin{pmatrix}
\left(\Delta X^2_{\vphi}\right)_{11} & \left(\Delta XP_{\vphi}\right)_{11}\\
\left(\Delta XP_{\vphi}\right)_{11} & \left(\Delta P^2_{\vphi}\right)_{11}
\end{pmatrix}.
\end{equation}
Our final step is to recover the variance of the quadrature $\hat{x}_\theta$. In order to do that, we introduce the $2\times2$ orthogonal and symplectic matrix
\begin{equation}
O_\theta =
\begin{pmatrix}
\cos\theta & \sin\theta\\
-\sin\theta & \cos\theta
\end{pmatrix},
\end{equation}
representing the action of a phase-shift $\e^{-\ii\theta}$, namely a clock-wise rotation of an angle $\theta$ in the first mode phase-space. The variance $\sigma_{\vphi}^2$ in~\eqref{eq:VarianceGeneric} is finally obtained by a direct computation
\begin{align}
\sigma^2_{\vphi}&=(O_\theta\Gamma'_{\vphi}O_\theta^\mathrm{T})_{11}\notag\\
&=\frac{1}{2} + P_{\vphi}(\cosh^2(r)+\cos(2f(\vphi)-2\theta)\cosh(r)\sinh(r)).
\end{align}
 
\section{Prior knowledge on the parameters}
\label{app:Prior}

In this appendix we will show that, with the setup presented in this work, a classical prior knowledge $\vphicl$, with an error $\delta\vphi = \vphi - \vphicl$ scaling as $1/\sqrt{N}$, is enough to correctly optimize our setup, satisfy condition~\eqref{eq:Condition1}, and thus ultimately reach Heisenberg scaling.

In general, in order to perform the optimization, the auxiliary stages $\hVi$ and $\hVo$ are chosen accordingly to a prior knowledge on the parameter, so that once the setup is correctly optimized, we can write $\hVi \equiv \hVi(\vphicl)$ and $\hVo \equiv \hVo(\vphicl)$. The one-photon transition probability $P_{\vphi}$ is by definition the square modulus of the (complex) scalar product of the two $M$-dimensional normalized vectors $\Uphi\Vi(\vphicl)\boldsymbol{e}_1$ and $\Vo^\dag(\vphicl)\boldsymbol{e}_1$, namely
\begin{equation}
P_{\vphi}=\abs{\boldsymbol{e}_1^\dagger \Vo(\vphicl) \Uphi \Vi(\vphicl)\boldsymbol{e}_1}^2 \equiv \eta(\vphi,\vphicl)
\end{equation}
with ${\boldsymbol{e}_1 = (1,0,\dots,0)^T}$, and $\eta$ is a smooth function of $\vphi$ and $\vphicl$ with global maxima along the condition $\vphicl=\vphi$ since, with a perfect prior knowledge of the parameters $\vphi$, the auxiliary stages are chosen so that $\abs{\boldsymbol{e}_1^\dagger \Vo(\vphi)\Uphi\Vi(\vphi)\boldsymbol{e}_1}=1$. If a small uncertainty $\delta\vphi=\vphi-\vphicl$ is present due to an imperfect prior knowledge on the parameter, then
\begin{align}
P_{\vphi} &= \eta(\vphi,\vphi-\delta\vphi) \notag \\
&=1 + \frac{1}{2}\sum_{i,j=1}^l (\partial^2_{i,j} \eta)_{\vphi} \delta\varphi_i \delta\varphi_j + \O(|\delta\vphi|^3),
\label{eq:AppendixP}
\end{align}
where $(\partial^2_{i,j} \eta)_{\vphi}$ is the second derivative of $\eta(\vphi,\cdot)$ with respect of the $i$-th and $j$-th components evaluated at $\vphi$. Comparing the expression of $P_{\vphi}$ in~\eqref{eq:AppendixP} with the condition~\eqref{eq:Condition1}, it is evident that the uncertainty allowed on the prior estimation, in order to correctly optimize $\hVo$ and ultimately reach Heisenberg scaling independently on the value of $\vphi$, must be of order $1/\sqrt{N}$, namely it must happen that $\delta\varphi_i = \O(N^{-1/2})$, for $i=1,\dots,l$. It is straightforward to notice that these results also hold if one of the two auxiliary gates is chosen independently from $\vphicl$, including the case of an identity operation corresponding to the absence of one gate.

\section{Derivation of the Fisher information matrix in~\eqref{eq:FisherMatrix}}\label{app:FisherMatrix}

In this appendix we will obtain the expression of the Fisher information matrix~\eqref{eq:FisherMatrix} from the general Fisher information matrix for a Gaussian distribution~\eqref{eq:FisherMatrixGauss} when condition~\eqref{eq:Condition1} and~\eqref{eq:Condition2} hold.

Since the dependence of $\sigma^2_{\vphi}$ in~\eqref{eq:VarianceGeneric} on the parameters $\vphi$ only appears through $P_{\vphi}$ and $f(\vphi)$, the gradient of the variance,
\begin{align}
\vnabla_{\vphi}\sigma^2_{\vphi} &=\left[ (\vnabla_{\vphi} P_{\vphi})\partial_P + (\vnabla_{\vphi} f(\vphi))\partial_\gamma\right] \sigma^2_{\vphi}\notag\\
&= \vnabla_{\vphi} P_{\vphi}\left(\sinh^2 r + \cos(2f(\vphi)-2\theta)\sinh r\cosh r\right) +\notag\\
&\quad -2P_{\vphi}\vnabla_{\vphi}f(\vphi)\sin(2f(\vphi)-2\theta)\sinh r \cosh r ,
\label{eq:VarianceGradient}
\end{align}
can be written in terms of $\vnabla_{\vphi} P_{\vphi}$ and $\vnabla_{\vphi}f(\vphi)$.
We now impose condition~\eqref{eq:Condition2} and fix $\theta = \theta_{\vphi}$, and evaluate the variance in~\eqref{eq:VarianceGeneric} and its gradient~\eqref{eq:VarianceGradient} in the large $N$ limit
\begin{align}
\sigma_{\vphi}^2 &= \frac{1-P_{\vphi}}{2} + P_{\vphi}\left(\frac{2k^2}{N}+\frac{1}{8N}\right) + \O\left(\frac{1}{N^2}\right),\\
\vnabla_{\vphi}\sigma^2_{\vphi} &= -\frac{1}{2}\vnabla_{\vphi}P_{\vphi}+4kP_{\vphi}\vnabla_{\vphi}f(\vphi) + \O\left(\frac{1}{N}\right).
\end{align}
Then, by imposing condition~\eqref{eq:Condition1} on $P_{\vphi}$, we get
\begin{align}
\sigma_{\vphi}^2 &= \left(2k^2+\frac{1}{8} + \frac{\ell}{2}\right)\frac{1}{N} + \O\left(\frac{1}{N^2}\right),\\
\vnabla_{\vphi}\sigma^2_{\vphi} &= 4k\vnabla_{\vphi}f(\vphi) + \O\left(\frac{1}{N}\right).
\end{align}
Therefore, the Fisher information matrix~\eqref{eq:FisherMatrixGauss} can be asymptotically written as
\begin{equation}
\I = 8\varrho(k,\ell) N^2 \bigl(\vnabla f(\vphi)\bigr) \bigl(\vnabla f(\vphi)\bigr)^T ,
\label{eq:C6}
\end{equation}
with
\begin{equation}
\varrho(k,\ell)=\left(\frac{8k}{16k^2 + 1 + 4\ell}\right)^2
\end{equation}
a positive and $N$-independent pre-factor.
\section{Proof of Heisenberg scaling}
\label{app:CRB}

In this appendix we will show that the Fisher information matrix $\I$ shown in~\eqref{eq:FisherMatrix} implies that the only functions of the parameter $\alpha(\vphi)$ which admit unbiased estimators with finite variance are of the form $\alpha(\vphi) \equiv g(f(\vphi))$, with $g(\cdot)$ a smooth function.

Since $\I$ in~\eqref{eq:C6} is a rank one matrix, it has a single non-zero eigenvalue $\lambda$. By direct calculation, the nonzero eigenvalue associated with the (normalized) eigenvector
\begin{equation}
\bm{v} = \frac{\vnabla f(\vphi)}{\abs{\vnabla f(\vphi)}}
\end{equation}
reads
\begin{equation}
\lambda = 8\varrho(k,\ell)\abs{\vnabla f(\vphi)}^2 N^2.
\end{equation}
In order for a given function $\alpha(\vphi)$ of the parameters to admit an unbiased estimator $\tilde{\alpha}$ with finite variance, it must happen that $\vnabla_{\vphi}\alpha(\vphi)$ belongs to the support of $\I$~\cite{Stoica2001,Gross2020}, which in our case is one dimensional and spanned by $\bm{v}$, whence
\begin{equation}
\vnabla_{\vphi}\alpha(\vphi)\propto \vnabla_{\vphi}f(\vphi),
\label{eq:alphacondition}
\end{equation}
which is verified for all $\vphi$, only if ${\alpha(\vphi) = g(f(\vphi))}$, with $g(\cdot)$ a differentiable real-valued function. For every unbiased estimator $\tilde{\alpha}\equiv\tilde{g}$ of such functions, the Cramér–Rao bound reads
\begin{equation}
\Var[\tilde{g}] \geq \left(\frac{\d g}{\d f}\right)^2\frac{1}{8\varrho(k,\ell)N^2},
\end{equation}
and, for the particular case $g(f(\vphi)) = f(\vphi)$
\begin{equation}
\Var[\tilde{f}] \geq \frac{1}{8\varrho(k,\ell)N^2}
\end{equation}

\section{Generalisation of the setup in Fig.~\ref{fig:Example2}}
\label{app:Generalisation}
In this appendix we will show that employing as local networks inside $\hUphi$ of Fig.~\ref{fig:Example2}, the network shown in Fig.~\ref{fig:Example1}, with conditions~\eqref{eq:OmegaDefinition} satisfied, still yields a setup which allows Heisenberg scaling sensitivity, this time for the estimation of a linear combination of \textit{functions} of parameters. In fact, even though these local networks do not exactly behave as single-mode phase-shifts, since~\eqref{eq:OmegaDefinition} holds locally, they still satisfy some local conditions
\begin{equation}
P_i\sim 1-\frac{\ell_i}{N},\qquad \ell_i\geq 0,\,i=1,\dots,m_2
\label{eq:Plocal}
\end{equation}
similar to the global condition~\eqref{eq:Condition1}, where $P_i$ in this case is the probability that a photon, injected in the first channel of the $i$-th local network, comes out from its upper channel. Thus, the probability amplitude~\eqref{eq:ProbAmpPhi} can generalizes to
\begin{align}
\chi_{\vphi}&=\e^{\ii L(\vphicl)}\sum_{i=1}^M \omega_i \sqrt{1-\frac{\ell_i}{N}} \e^{\ii \delta\varphi_i}=\notag\\
&=\e^{\ii L(\vphicl)}\left(1+\sum_{i=1}^M \ii \omega_i\delta\varphi_i - \frac{1}{2} \sum_{i=1}^M \omega_i\left(\delta\varphi_i^2+\frac{\ell_i}{N}\right)\right)\notag\\
& + \O(N^{-3/2}),
\label{eq:ProbAmpPhiApp}
\end{align}
where we made use of condition~\eqref{eq:Plocal} to write the transition amplitudes associated with each channel of $\hUphi$. Exploiting once again the requirement that $\delta\varphi_i=\O(N^{-1/2})$, for $i=1,\dots,M$, we notice that the one-photon probability,
\begin{align}
P_{\vphi}&=\abs{\e^{\ii L(\vphicl)}\sum_{i=1}^M \omega_i \sqrt{1-\frac{\ell_i}{N}} \e^{\ii \delta\varphi_i}}^2=\notag\\
&= 1 + \left(\sum_{i=1}^M \omega_i\delta\varphi_i\right)^2 - \sum_{i=1}^M \omega_i\left(\delta\varphi_i^2+\frac{\ell_i}{N}\right) + \O(N^{-3/2})\notag\\
&\equiv 1 - \frac{\ell}{N}+ \O(N^{-3/2})
\end{align}
still satisfies condition~\eqref{eq:Condition1}, so that Heisenberg scaling sensitivity in the estimation of the total acquired phase shown in~\eqref{eq:gammaExample1} is achieved.

\nocite{*}
\bibliography{references}

\begin{thebibliography}{35}%
\makeatletter
\providecommand \@ifxundefined [1]{%
 \@ifx{#1\undefined}
}%
\providecommand \@ifnum [1]{%
 \ifnum #1\expandafter \@firstoftwo
 \else \expandafter \@secondoftwo
 \fi
}%
\providecommand \@ifx [1]{%
 \ifx #1\expandafter \@firstoftwo
 \else \expandafter \@secondoftwo
 \fi
}%
\providecommand \natexlab [1]{#1}%
\providecommand \enquote  [1]{``#1''}%
\providecommand \bibnamefont  [1]{#1}%
\providecommand \bibfnamefont [1]{#1}%
\providecommand \citenamefont [1]{#1}%
\providecommand \href@noop [0]{\@secondoftwo}%
\providecommand \href [0]{\begingroup \@sanitize@url \@href}%
\providecommand \@href[1]{\@@startlink{#1}\@@href}%
\providecommand \@@href[1]{\endgroup#1\@@endlink}%
\providecommand \@sanitize@url [0]{\catcode `\\12\catcode `\$12\catcode
  `\&12\catcode `\#12\catcode `\^12\catcode `\_12\catcode `\%12\relax}%
\providecommand \@@startlink[1]{}%
\providecommand \@@endlink[0]{}%
\providecommand \url  [0]{\begingroup\@sanitize@url \@url }%
\providecommand \@url [1]{\endgroup\@href {#1}{\urlprefix }}%
\providecommand \urlprefix  [0]{URL }%
\providecommand \Eprint [0]{\href }%
\providecommand \doibase [0]{http://dx.doi.org/}%
\providecommand \selectlanguage [0]{\@gobble}%
\providecommand \bibinfo  [0]{\@secondoftwo}%
\providecommand \bibfield  [0]{\@secondoftwo}%
\providecommand \translation [1]{[#1]}%
\providecommand \BibitemOpen [0]{}%
\providecommand \bibitemStop [0]{}%
\providecommand \bibitemNoStop [0]{.\EOS\space}%
\providecommand \EOS [0]{\spacefactor3000\relax}%
\providecommand \BibitemShut  [1]{\csname bibitem#1\endcsname}%
\let\auto@bib@innerbib\@empty
\bibitem [{\citenamefont {Giovannetti}\ \emph {et~al.}(2004)\citenamefont
  {Giovannetti}, \citenamefont {Lloyd},\ and\ \citenamefont
  {Maccone}}]{Giovannetti2004}%
  \BibitemOpen
  \bibfield  {author} {\bibinfo {author} {\bibfnamefont {Vittorio}\
  \bibnamefont {Giovannetti}}, \bibinfo {author} {\bibfnamefont {Seth}\
  \bibnamefont {Lloyd}}, \ and\ \bibinfo {author} {\bibfnamefont {Lorenzo}\
  \bibnamefont {Maccone}},\ }\bibfield  {title} {\enquote {\bibinfo {title}
  {Quantum-enhanced measurements: Beating the standard quantum limit},}\ }\href
  {\doibase 10.1126/science.1104149} {\bibfield  {journal} {\bibinfo  {journal}
  {Science}\ }\textbf {\bibinfo {volume} {306}},\ \bibinfo {pages} {1330--1336}
  (\bibinfo {year} {2004})}\BibitemShut {NoStop}%
\bibitem [{\citenamefont {Giovannetti}\ \emph {et~al.}(2006)\citenamefont
  {Giovannetti}, \citenamefont {Lloyd},\ and\ \citenamefont
  {Maccone}}]{Giovannetti2006}%
  \BibitemOpen
  \bibfield  {author} {\bibinfo {author} {\bibfnamefont {Vittorio}\
  \bibnamefont {Giovannetti}}, \bibinfo {author} {\bibfnamefont {Seth}\
  \bibnamefont {Lloyd}}, \ and\ \bibinfo {author} {\bibfnamefont {Lorenzo}\
  \bibnamefont {Maccone}},\ }\bibfield  {title} {\enquote {\bibinfo {title}
  {Quantum metrology},}\ }\href {\doibase 10.1103/PhysRevLett.96.010401}
  {\bibfield  {journal} {\bibinfo  {journal} {Phys. Rev. Lett.}\ }\textbf
  {\bibinfo {volume} {96}},\ \bibinfo {pages} {010401} (\bibinfo {year}
  {2006})}\BibitemShut {NoStop}%
\bibitem [{\citenamefont {Dowling}(2008)}]{Dowling2008}%
  \BibitemOpen
  \bibfield  {author} {\bibinfo {author} {\bibfnamefont {Jonathan~P.}\
  \bibnamefont {Dowling}},\ }\bibfield  {title} {\enquote {\bibinfo {title}
  {Quantum optical metrology -- the lowdown on high-n00n states},}\ }\href
  {\doibase 10.1080/00107510802091298} {\bibfield  {journal} {\bibinfo
  {journal} {Contemporary Physics}\ }\textbf {\bibinfo {volume} {49}},\
  \bibinfo {pages} {125--143} (\bibinfo {year} {2008})}\BibitemShut {NoStop}%
\bibitem [{\citenamefont {Giovannetti}\ \emph {et~al.}(2011)\citenamefont
  {Giovannetti}, \citenamefont {Lloyd},\ and\ \citenamefont
  {Maccone}}]{Giovannetti2011}%
  \BibitemOpen
  \bibfield  {author} {\bibinfo {author} {\bibfnamefont {Vittorio}\
  \bibnamefont {Giovannetti}}, \bibinfo {author} {\bibfnamefont {Seth}\
  \bibnamefont {Lloyd}}, \ and\ \bibinfo {author} {\bibfnamefont {Lorenzo}\
  \bibnamefont {Maccone}},\ }\bibfield  {title} {\enquote {\bibinfo {title}
  {Advances in quantum metrology},}\ }\href {\doibase 10.1038/nphoton.2011.35}
  {\bibfield  {journal} {\bibinfo  {journal} {Nature Photonics}\ }\textbf
  {\bibinfo {volume} {5}},\ \bibinfo {pages} {010401} (\bibinfo {year}
  {2011})}\BibitemShut {NoStop}%
\bibitem [{\citenamefont {{Dowling}}\ and\ \citenamefont
  {{Seshadreesan}}(2015)}]{Dowling2015}%
  \BibitemOpen
  \bibfield  {author} {\bibinfo {author} {\bibfnamefont {J.~P.}\ \bibnamefont
  {{Dowling}}}\ and\ \bibinfo {author} {\bibfnamefont {K.~P.}\ \bibnamefont
  {{Seshadreesan}}},\ }\bibfield  {title} {\enquote {\bibinfo {title} {Quantum
  optical technologies for metrology, sensing, and imaging},}\ }\href {\doibase
  10.1109/JLT.2014.2386795} {\bibfield  {journal} {\bibinfo  {journal} {Journal
  of Lightwave Technology}\ }\textbf {\bibinfo {volume} {33}},\ \bibinfo
  {pages} {2359--2370} (\bibinfo {year} {2015})}\BibitemShut {NoStop}%
\bibitem [{\citenamefont {De~Pasquale}\ \emph {et~al.}(2015)\citenamefont
  {De~Pasquale}, \citenamefont {Facchi}, \citenamefont {Florio}, \citenamefont
  {Giovannetti}, \citenamefont {Matsuoka},\ and\ \citenamefont
  {Yuasa}}]{DePasquale2015}%
  \BibitemOpen
  \bibfield  {author} {\bibinfo {author} {\bibfnamefont {Antonella}\
  \bibnamefont {De~Pasquale}}, \bibinfo {author} {\bibfnamefont {Paolo}\
  \bibnamefont {Facchi}}, \bibinfo {author} {\bibfnamefont {Giuseppe}\
  \bibnamefont {Florio}}, \bibinfo {author} {\bibfnamefont {Vittorio}\
  \bibnamefont {Giovannetti}}, \bibinfo {author} {\bibfnamefont {Koji}\
  \bibnamefont {Matsuoka}}, \ and\ \bibinfo {author} {\bibfnamefont {Kazuya}\
  \bibnamefont {Yuasa}},\ }\bibfield  {title} {\enquote {\bibinfo {title}
  {Two-mode bosonic quantum metrology with number fluctuations},}\ }\href
  {\doibase 10.1103/physreva.92.042115} {\bibfield  {journal} {\bibinfo
  {journal} {Physical Review A}\ }\textbf {\bibinfo {volume} {92}} (\bibinfo
  {year} {2015}),\ 10.1103/physreva.92.042115}\BibitemShut {NoStop}%
\bibitem [{\citenamefont {Zhou}\ \emph {et~al.}(2018)\citenamefont {Zhou},
  \citenamefont {Zhang}, \citenamefont {Preskill},\ and\ \citenamefont
  {Jiang}}]{Zhou2018}%
  \BibitemOpen
  \bibfield  {author} {\bibinfo {author} {\bibfnamefont {Sisi}\ \bibnamefont
  {Zhou}}, \bibinfo {author} {\bibfnamefont {Mengzhen}\ \bibnamefont {Zhang}},
  \bibinfo {author} {\bibfnamefont {John}\ \bibnamefont {Preskill}}, \ and\
  \bibinfo {author} {\bibfnamefont {Liang}\ \bibnamefont {Jiang}},\ }\bibfield
  {title} {\enquote {\bibinfo {title} {Achieving the heisenberg limit in
  quantum metrology using quantum error correction},}\ }\href {\doibase
  10.1038/s41467-017-02510-3} {\bibfield  {journal} {\bibinfo  {journal}
  {Nature Communications}\ }\textbf {\bibinfo {volume} {9}} (\bibinfo {year}
  {2018}),\ 10.1038/s41467-017-02510-3}\BibitemShut {NoStop}%
\bibitem [{\citenamefont {Ge}\ \emph {et~al.}(2018)\citenamefont {Ge},
  \citenamefont {Jacobs}, \citenamefont {Eldredge}, \citenamefont {Gorshkov},\
  and\ \citenamefont {Foss-Feig}}]{Ge2018}%
  \BibitemOpen
  \bibfield  {author} {\bibinfo {author} {\bibfnamefont {Wenchao}\ \bibnamefont
  {Ge}}, \bibinfo {author} {\bibfnamefont {Kurt}\ \bibnamefont {Jacobs}},
  \bibinfo {author} {\bibfnamefont {Zachary}\ \bibnamefont {Eldredge}},
  \bibinfo {author} {\bibfnamefont {Alexey~V.}\ \bibnamefont {Gorshkov}}, \
  and\ \bibinfo {author} {\bibfnamefont {Michael}\ \bibnamefont {Foss-Feig}},\
  }\bibfield  {title} {\enquote {\bibinfo {title} {Distributed quantum
  metrology with linear networks and separable inputs},}\ }\href {\doibase
  10.1103/PhysRevLett.121.043604} {\bibfield  {journal} {\bibinfo  {journal}
  {Phys. Rev. Lett.}\ }\textbf {\bibinfo {volume} {121}},\ \bibinfo {pages}
  {043604} (\bibinfo {year} {2018})}\BibitemShut {NoStop}%
\bibitem [{\citenamefont {Qian}\ \emph {et~al.}(2019)\citenamefont {Qian},
  \citenamefont {Eldredge}, \citenamefont {Ge}, \citenamefont {Pagano},
  \citenamefont {Monroe}, \citenamefont {Porto},\ and\ \citenamefont
  {Gorshkov}}]{Qian2019}%
  \BibitemOpen
  \bibfield  {author} {\bibinfo {author} {\bibfnamefont {Kevin}\ \bibnamefont
  {Qian}}, \bibinfo {author} {\bibfnamefont {Zachary}\ \bibnamefont
  {Eldredge}}, \bibinfo {author} {\bibfnamefont {Wenchao}\ \bibnamefont {Ge}},
  \bibinfo {author} {\bibfnamefont {Guido}\ \bibnamefont {Pagano}}, \bibinfo
  {author} {\bibfnamefont {Christopher}\ \bibnamefont {Monroe}}, \bibinfo
  {author} {\bibfnamefont {J.~V.}\ \bibnamefont {Porto}}, \ and\ \bibinfo
  {author} {\bibfnamefont {Alexey~V.}\ \bibnamefont {Gorshkov}},\ }\bibfield
  {title} {\enquote {\bibinfo {title} {Heisenberg-scaling measurement protocol
  for analytic functions with quantum sensor networks},}\ }\href {\doibase
  10.1103/PhysRevA.100.042304} {\bibfield  {journal} {\bibinfo  {journal}
  {Phys. Rev. A}\ }\textbf {\bibinfo {volume} {100}},\ \bibinfo {pages}
  {042304} (\bibinfo {year} {2019})}\BibitemShut {NoStop}%
\bibitem [{\citenamefont {McConnell}\ \emph {et~al.}(2017)\citenamefont
  {McConnell}, \citenamefont {Low}, \citenamefont {Yoder}, \citenamefont
  {Bruzewicz}, \citenamefont {Chuang}, \citenamefont {Chiaverini},\ and\
  \citenamefont {Sage}}]{McConnell2017}%
  \BibitemOpen
  \bibfield  {author} {\bibinfo {author} {\bibfnamefont {Robert}\ \bibnamefont
  {McConnell}}, \bibinfo {author} {\bibfnamefont {Guang~Hao}\ \bibnamefont
  {Low}}, \bibinfo {author} {\bibfnamefont {Theodore~J.}\ \bibnamefont
  {Yoder}}, \bibinfo {author} {\bibfnamefont {Colin~D.}\ \bibnamefont
  {Bruzewicz}}, \bibinfo {author} {\bibfnamefont {Isaac~L.}\ \bibnamefont
  {Chuang}}, \bibinfo {author} {\bibfnamefont {John}\ \bibnamefont
  {Chiaverini}}, \ and\ \bibinfo {author} {\bibfnamefont {Jeremy~M.}\
  \bibnamefont {Sage}},\ }\bibfield  {title} {\enquote {\bibinfo {title}
  {Heisenberg scaling of imaging resolution by coherent enhancement},}\ }\href
  {\doibase 10.1103/PhysRevA.96.051801} {\bibfield  {journal} {\bibinfo
  {journal} {Phys. Rev. A}\ }\textbf {\bibinfo {volume} {96}},\ \bibinfo
  {pages} {051801} (\bibinfo {year} {2017})}\BibitemShut {NoStop}%
\bibitem [{\citenamefont {Untern\"{a}hrer}\ \emph {et~al.}(2018)\citenamefont
  {Untern\"{a}hrer}, \citenamefont {Bessire}, \citenamefont {Gasparini},
  \citenamefont {Perenzoni},\ and\ \citenamefont {Stefanov}}]{Unternahrer2018}%
  \BibitemOpen
  \bibfield  {author} {\bibinfo {author} {\bibfnamefont {Manuel}\ \bibnamefont
  {Untern\"{a}hrer}}, \bibinfo {author} {\bibfnamefont {B\"{a}nz}\ \bibnamefont
  {Bessire}}, \bibinfo {author} {\bibfnamefont {Leonardo}\ \bibnamefont
  {Gasparini}}, \bibinfo {author} {\bibfnamefont {Matteo}\ \bibnamefont
  {Perenzoni}}, \ and\ \bibinfo {author} {\bibfnamefont {Andr\'{e}}\
  \bibnamefont {Stefanov}},\ }\bibfield  {title} {\enquote {\bibinfo {title}
  {Super-resolution quantum imaging at the heisenberg limit},}\ }\href
  {\doibase 10.1364/OPTICA.5.001150} {\bibfield  {journal} {\bibinfo  {journal}
  {Optica}\ }\textbf {\bibinfo {volume} {5}},\ \bibinfo {pages} {1150--1154}
  (\bibinfo {year} {2018})}\BibitemShut {NoStop}%
\bibitem [{\citenamefont {De~Pasquale}\ and\ \citenamefont
  {Stace}(2018)}]{DePasquale2018}%
  \BibitemOpen
  \bibfield  {author} {\bibinfo {author} {\bibfnamefont {Antonella}\
  \bibnamefont {De~Pasquale}}\ and\ \bibinfo {author} {\bibfnamefont
  {Thomas~M.}\ \bibnamefont {Stace}},\ }\bibfield  {title} {\enquote {\bibinfo
  {title} {Quantum thermometry},}\ }\href {\doibase
  10.1007/978-3-319-99046-0_21} {\bibfield  {journal} {\bibinfo  {journal}
  {Thermodynamics in the Quantum Regime}\ ,\ \bibinfo {pages} {503–527}}
  (\bibinfo {year} {2018})}\BibitemShut {NoStop}%
\bibitem [{\citenamefont {Seah}\ \emph {et~al.}(2019)\citenamefont {Seah},
  \citenamefont {Nimmrichter}, \citenamefont {Grimmer}, \citenamefont {Santos},
  \citenamefont {Scarani},\ and\ \citenamefont {Landi}}]{Seah2019}%
  \BibitemOpen
  \bibfield  {author} {\bibinfo {author} {\bibfnamefont {Stella}\ \bibnamefont
  {Seah}}, \bibinfo {author} {\bibfnamefont {Stefan}\ \bibnamefont
  {Nimmrichter}}, \bibinfo {author} {\bibfnamefont {Daniel}\ \bibnamefont
  {Grimmer}}, \bibinfo {author} {\bibfnamefont {Jader~P.}\ \bibnamefont
  {Santos}}, \bibinfo {author} {\bibfnamefont {Valerio}\ \bibnamefont
  {Scarani}}, \ and\ \bibinfo {author} {\bibfnamefont {Gabriel~T.}\
  \bibnamefont {Landi}},\ }\bibfield  {title} {\enquote {\bibinfo {title}
  {Collisional quantum thermometry},}\ }\href {\doibase
  10.1103/PhysRevLett.123.180602} {\bibfield  {journal} {\bibinfo  {journal}
  {Phys. Rev. Lett.}\ }\textbf {\bibinfo {volume} {123}},\ \bibinfo {pages}
  {180602} (\bibinfo {year} {2019})}\BibitemShut {NoStop}%
\bibitem [{\citenamefont {Razzoli}\ \emph {et~al.}(2019)\citenamefont
  {Razzoli}, \citenamefont {Ghirardi}, \citenamefont {Siloi}, \citenamefont
  {Bordone},\ and\ \citenamefont {Paris}}]{Razzoli2019}%
  \BibitemOpen
  \bibfield  {author} {\bibinfo {author} {\bibfnamefont {Luca}\ \bibnamefont
  {Razzoli}}, \bibinfo {author} {\bibfnamefont {Luca}\ \bibnamefont
  {Ghirardi}}, \bibinfo {author} {\bibfnamefont {Ilaria}\ \bibnamefont
  {Siloi}}, \bibinfo {author} {\bibfnamefont {Paolo}\ \bibnamefont {Bordone}},
  \ and\ \bibinfo {author} {\bibfnamefont {Matteo G.~A.}\ \bibnamefont
  {Paris}},\ }\bibfield  {title} {\enquote {\bibinfo {title} {Lattice quantum
  magnetometry},}\ }\href {\doibase 10.1103/PhysRevA.99.062330} {\bibfield
  {journal} {\bibinfo  {journal} {Phys. Rev. A}\ }\textbf {\bibinfo {volume}
  {99}},\ \bibinfo {pages} {062330} (\bibinfo {year} {2019})}\BibitemShut
  {NoStop}%
\bibitem [{\citenamefont {Bhattacharjee}\ \emph {et~al.}(2020)\citenamefont
  {Bhattacharjee}, \citenamefont {Bhattacharya}, \citenamefont {Niedenzu},
  \citenamefont {Mukherjee},\ and\ \citenamefont {Dutta}}]{Bhattacharjee2020}%
  \BibitemOpen
  \bibfield  {author} {\bibinfo {author} {\bibfnamefont {Sourav}\ \bibnamefont
  {Bhattacharjee}}, \bibinfo {author} {\bibfnamefont {Utso}\ \bibnamefont
  {Bhattacharya}}, \bibinfo {author} {\bibfnamefont {Wolfgang}\ \bibnamefont
  {Niedenzu}}, \bibinfo {author} {\bibfnamefont {Victor}\ \bibnamefont
  {Mukherjee}}, \ and\ \bibinfo {author} {\bibfnamefont {Amit}\ \bibnamefont
  {Dutta}},\ }\bibfield  {title} {\enquote {\bibinfo {title} {Quantum
  magnetometry using two-stroke thermal machines},}\ }\href@noop {} {\bibfield
  {journal} {\bibinfo  {journal} {New Journal of Physics}\ }\textbf {\bibinfo
  {volume} {22}},\ \bibinfo {pages} {013024} (\bibinfo {year}
  {2020})}\BibitemShut {NoStop}%
\bibitem [{\citenamefont {Aasi}\ \emph {et~al.}(2013)\citenamefont {Aasi},
  \citenamefont {Abadie}, \citenamefont {Abbott}, \citenamefont {Abbott},
  \citenamefont {Abbott}, \citenamefont {Abernathy}, \citenamefont {Adams},
  \citenamefont {Adams}, \citenamefont {Addesso}, \citenamefont {Adhikari},\
  and\ \citenamefont {et~al.}}]{Ligo2013}%
  \BibitemOpen
  \bibfield  {author} {\bibinfo {author} {\bibfnamefont {J.}~\bibnamefont
  {Aasi}}, \bibinfo {author} {\bibfnamefont {J.}~\bibnamefont {Abadie}},
  \bibinfo {author} {\bibfnamefont {B.~P.}\ \bibnamefont {Abbott}}, \bibinfo
  {author} {\bibfnamefont {R.}~\bibnamefont {Abbott}}, \bibinfo {author}
  {\bibfnamefont {T.~D.}\ \bibnamefont {Abbott}}, \bibinfo {author}
  {\bibfnamefont {M.~R.}\ \bibnamefont {Abernathy}}, \bibinfo {author}
  {\bibfnamefont {C.}~\bibnamefont {Adams}}, \bibinfo {author} {\bibfnamefont
  {T.}~\bibnamefont {Adams}}, \bibinfo {author} {\bibfnamefont
  {P.}~\bibnamefont {Addesso}}, \bibinfo {author} {\bibfnamefont {R.~X.}\
  \bibnamefont {Adhikari}}, \ and\ \bibinfo {author} {\bibnamefont {et~al.}},\
  }\bibfield  {title} {\enquote {\bibinfo {title} {Enhanced sensitivity of the
  ligo gravitational wave detector by using squeezed states of light},}\ }\href
  {\doibase 10.1038/nphoton.2013.177} {\bibfield  {journal} {\bibinfo
  {journal} {Nature Photonics}\ }\textbf {\bibinfo {volume} {7}},\ \bibinfo
  {pages} {613–619} (\bibinfo {year} {2013})}\BibitemShut {NoStop}%
\bibitem [{\citenamefont {Monras}(2006)}]{Monras2006}%
  \BibitemOpen
  \bibfield  {author} {\bibinfo {author} {\bibfnamefont {Alex}\ \bibnamefont
  {Monras}},\ }\bibfield  {title} {\enquote {\bibinfo {title} {Optimal phase
  measurements with pure gaussian states},}\ }\href {\doibase
  10.1103/PhysRevA.73.033821} {\bibfield  {journal} {\bibinfo  {journal} {Phys.
  Rev. A}\ }\textbf {\bibinfo {volume} {73}},\ \bibinfo {pages} {033821}
  (\bibinfo {year} {2006})}\BibitemShut {NoStop}%
\bibitem [{\citenamefont {Pezz\'e}\ and\ \citenamefont
  {Smerzi}(2008)}]{Pezze2008}%
  \BibitemOpen
  \bibfield  {author} {\bibinfo {author} {\bibfnamefont {Luca}\ \bibnamefont
  {Pezz\'e}}\ and\ \bibinfo {author} {\bibfnamefont {Augusto}\ \bibnamefont
  {Smerzi}},\ }\bibfield  {title} {\enquote {\bibinfo {title} {Mach-zehnder
  interferometry at the heisenberg limit with coherent and squeezed-vacuum
  light},}\ }\href {\doibase 10.1103/PhysRevLett.100.073601} {\bibfield
  {journal} {\bibinfo  {journal} {Phys. Rev. Lett.}\ }\textbf {\bibinfo
  {volume} {100}},\ \bibinfo {pages} {073601} (\bibinfo {year}
  {2008})}\BibitemShut {NoStop}%
\bibitem [{\citenamefont {Aspachs}\ \emph {et~al.}(2009)\citenamefont
  {Aspachs}, \citenamefont {Calsamiglia}, \citenamefont {Mu\~noz Tapia},\ and\
  \citenamefont {Bagan}}]{Aspachs2009}%
  \BibitemOpen
  \bibfield  {author} {\bibinfo {author} {\bibfnamefont {M.}~\bibnamefont
  {Aspachs}}, \bibinfo {author} {\bibfnamefont {J.}~\bibnamefont
  {Calsamiglia}}, \bibinfo {author} {\bibfnamefont {R.}~\bibnamefont {Mu\~noz
  Tapia}}, \ and\ \bibinfo {author} {\bibfnamefont {E.}~\bibnamefont {Bagan}},\
  }\bibfield  {title} {\enquote {\bibinfo {title} {Phase estimation for thermal
  gaussian states},}\ }\href {\doibase 10.1103/PhysRevA.79.033834} {\bibfield
  {journal} {\bibinfo  {journal} {Phys. Rev. A}\ }\textbf {\bibinfo {volume}
  {79}},\ \bibinfo {pages} {033834} (\bibinfo {year} {2009})}\BibitemShut
  {NoStop}%
\bibitem [{\citenamefont {Lang}\ and\ \citenamefont {Caves}(2013)}]{Lang2013}%
  \BibitemOpen
  \bibfield  {author} {\bibinfo {author} {\bibfnamefont {Matthias~D.}\
  \bibnamefont {Lang}}\ and\ \bibinfo {author} {\bibfnamefont {Carlton~M.}\
  \bibnamefont {Caves}},\ }\bibfield  {title} {\enquote {\bibinfo {title}
  {Optimal quantum-enhanced interferometry using a laser power source},}\
  }\href {\doibase 10.1103/PhysRevLett.111.173601} {\bibfield  {journal}
  {\bibinfo  {journal} {Phys. Rev. Lett.}\ }\textbf {\bibinfo {volume} {111}},\
  \bibinfo {pages} {173601} (\bibinfo {year} {2013})}\BibitemShut {NoStop}%
\bibitem [{\citenamefont {{Maccone}}\ and\ \citenamefont
  {{Riccardi}}(2019)}]{Maccone2019}%
  \BibitemOpen
  \bibfield  {author} {\bibinfo {author} {\bibfnamefont {Lorenzo}\ \bibnamefont
  {{Maccone}}}\ and\ \bibinfo {author} {\bibfnamefont {Alberto}\ \bibnamefont
  {{Riccardi}}},\ }\bibfield  {title} {\enquote {\bibinfo {title} {{Squeezing
  metrology}},}\ }\href@noop {} {\ ,\ \bibinfo {pages} {arXiv:1901.07482}
  (\bibinfo {year} {2019})},\ \Eprint {http://arxiv.org/abs/1901.07482}
  {arXiv:1901.07482 [quant-ph]} \BibitemShut {NoStop}%
\bibitem [{\citenamefont {Matsubara}\ \emph {et~al.}(2019)\citenamefont
  {Matsubara}, \citenamefont {Facchi}, \citenamefont {Giovannetti},\ and\
  \citenamefont {Yuasa}}]{Matsubara2019}%
  \BibitemOpen
  \bibfield  {author} {\bibinfo {author} {\bibfnamefont {Teruo}\ \bibnamefont
  {Matsubara}}, \bibinfo {author} {\bibfnamefont {Paolo}\ \bibnamefont
  {Facchi}}, \bibinfo {author} {\bibfnamefont {Vittorio}\ \bibnamefont
  {Giovannetti}}, \ and\ \bibinfo {author} {\bibfnamefont {Kazuya}\
  \bibnamefont {Yuasa}},\ }\bibfield  {title} {\enquote {\bibinfo {title}
  {Optimal gaussian metrology for generic multimode interferometric circuit},}\
  }\href {\doibase 10.1088/1367-2630/ab0604} {\bibfield  {journal} {\bibinfo
  {journal} {New Journal of Physics}\ }\textbf {\bibinfo {volume} {21}},\
  \bibinfo {pages} {033014} (\bibinfo {year} {2019})}\BibitemShut {NoStop}%
\bibitem [{\citenamefont {Oh}\ \emph {et~al.}(2019)\citenamefont {Oh},
  \citenamefont {Lee}, \citenamefont {Rockstuhl}, \citenamefont {Jeong},
  \citenamefont {Kim}, \citenamefont {Nha},\ and\ \citenamefont
  {Lee}}]{Oh2019}%
  \BibitemOpen
  \bibfield  {author} {\bibinfo {author} {\bibfnamefont {Changhun}\
  \bibnamefont {Oh}}, \bibinfo {author} {\bibfnamefont {Changhyoup}\
  \bibnamefont {Lee}}, \bibinfo {author} {\bibfnamefont {Carsten}\ \bibnamefont
  {Rockstuhl}}, \bibinfo {author} {\bibfnamefont {Hyunseok}\ \bibnamefont
  {Jeong}}, \bibinfo {author} {\bibfnamefont {Jaewan}\ \bibnamefont {Kim}},
  \bibinfo {author} {\bibfnamefont {Hyunchul}\ \bibnamefont {Nha}}, \ and\
  \bibinfo {author} {\bibfnamefont {Su-Yong}\ \bibnamefont {Lee}},\ }\bibfield
  {title} {\enquote {\bibinfo {title} {Optimal gaussian measurements for phase
  estimation in single-mode gaussian metrology},}\ }\href {\doibase
  10.1038/s41534-019-0124-4} {\bibfield  {journal} {\bibinfo  {journal} {npj
  Quantum Information}\ }\textbf {\bibinfo {volume} {5}},\ \bibinfo {pages}
  {10} (\bibinfo {year} {2019})}\BibitemShut {NoStop}%
\bibitem [{\citenamefont {Gatto}\ \emph {et~al.}(2019)\citenamefont {Gatto},
  \citenamefont {Facchi}, \citenamefont {Narducci},\ and\ \citenamefont
  {Tamma}}]{Gatto2019}%
  \BibitemOpen
  \bibfield  {author} {\bibinfo {author} {\bibfnamefont {Dario}\ \bibnamefont
  {Gatto}}, \bibinfo {author} {\bibfnamefont {Paolo}\ \bibnamefont {Facchi}},
  \bibinfo {author} {\bibfnamefont {Frank~A.}\ \bibnamefont {Narducci}}, \ and\
  \bibinfo {author} {\bibfnamefont {Vincenzo}\ \bibnamefont {Tamma}},\
  }\bibfield  {title} {\enquote {\bibinfo {title} {Distributed quantum
  metrology with a single squeezed-vacuum source},}\ }\href {\doibase
  10.1103/PhysRevResearch.1.032024} {\bibfield  {journal} {\bibinfo  {journal}
  {Phys. Rev. Research}\ }\textbf {\bibinfo {volume} {1}},\ \bibinfo {pages}
  {032024} (\bibinfo {year} {2019})}\BibitemShut {NoStop}%
\bibitem [{\citenamefont {Gatto}\ \emph {et~al.}(0)\citenamefont {Gatto},
  \citenamefont {Facchi},\ and\ \citenamefont {Tamma}}]{Gatto2020}%
  \BibitemOpen
  \bibfield  {author} {\bibinfo {author} {\bibfnamefont {Dario}\ \bibnamefont
  {Gatto}}, \bibinfo {author} {\bibfnamefont {Paolo}\ \bibnamefont {Facchi}}, \
  and\ \bibinfo {author} {\bibfnamefont {Vincenzo}\ \bibnamefont {Tamma}},\
  }\bibfield  {title} {\enquote {\bibinfo {title} {Phase space
  heisenberg-limited estimation of the average phase shift in a mach–zehnder
  interferometer},}\ }\href {\doibase 10.1142/S0219749919410193} {\bibfield
  {journal} {\bibinfo  {journal} {International Journal of Quantum
  Information}\ }\textbf {\bibinfo {volume} {0}},\ \bibinfo {pages} {1941019}
  (\bibinfo {year} {0})},\ \Eprint
  {http://arxiv.org/abs/https://doi.org/10.1142/S0219749919410193}
  {https://doi.org/10.1142/S0219749919410193} \BibitemShut {NoStop}%
\bibitem [{\citenamefont {Gramegna}\ \emph {et~al.}(2021)\citenamefont
  {Gramegna}, \citenamefont {Triggiani}, \citenamefont {Facchi}, \citenamefont
  {Narducci},\ and\ \citenamefont {Tamma}}]{Gramegna2020Typicality}%
  \BibitemOpen
  \bibfield  {author} {\bibinfo {author} {\bibfnamefont {Giovanni}\
  \bibnamefont {Gramegna}}, \bibinfo {author} {\bibfnamefont {Danilo}\
  \bibnamefont {Triggiani}}, \bibinfo {author} {\bibfnamefont {Paolo}\
  \bibnamefont {Facchi}}, \bibinfo {author} {\bibfnamefont {Frank~A.}\
  \bibnamefont {Narducci}}, \ and\ \bibinfo {author} {\bibfnamefont {Vincenzo}\
  \bibnamefont {Tamma}},\ }\bibfield  {title} {\enquote {\bibinfo {title}
  {Typicality of heisenberg scaling precision in multimode quantum
  metrology},}\ }\href {\doibase 10.1103/PhysRevResearch.3.013152} {\bibfield
  {journal} {\bibinfo  {journal} {Phys. Rev. Research}\ }\textbf {\bibinfo
  {volume} {3}},\ \bibinfo {pages} {013152} (\bibinfo {year}
  {2021})}\BibitemShut {NoStop}%
\bibitem [{\citenamefont {Gramegna}\ \emph {et~al.}(2020)\citenamefont
  {Gramegna}, \citenamefont {Triggiani}, \citenamefont {Facchi}, \citenamefont
  {Narducci},\ and\ \citenamefont {Tamma}}]{Gramegna2020Letter}%
  \BibitemOpen
  \bibfield  {author} {\bibinfo {author} {\bibfnamefont {Giovanni}\
  \bibnamefont {Gramegna}}, \bibinfo {author} {\bibfnamefont {Danilo}\
  \bibnamefont {Triggiani}}, \bibinfo {author} {\bibfnamefont {Paolo}\
  \bibnamefont {Facchi}}, \bibinfo {author} {\bibfnamefont {Frank~A.}\
  \bibnamefont {Narducci}}, \ and\ \bibinfo {author} {\bibfnamefont {Vincenzo}\
  \bibnamefont {Tamma}},\ }\bibfield  {title} {\enquote {\bibinfo {title}
  {Heisenberg scaling precision in multi-mode distributed quantum metrology},}\
  }\href@noop {} {\  (\bibinfo {year} {2020})},\ \Eprint
  {http://arxiv.org/abs/2003.12550} {arXiv:2003.12550 [quant-ph]} \BibitemShut
  {NoStop}%
\bibitem [{\citenamefont {Zhuang}\ \emph {et~al.}(2018)\citenamefont {Zhuang},
  \citenamefont {Zhang},\ and\ \citenamefont {Shapiro}}]{Zhuang2018}%
  \BibitemOpen
  \bibfield  {author} {\bibinfo {author} {\bibfnamefont {Quntao}\ \bibnamefont
  {Zhuang}}, \bibinfo {author} {\bibfnamefont {Zheshen}\ \bibnamefont {Zhang}},
  \ and\ \bibinfo {author} {\bibfnamefont {Jeffrey~H.}\ \bibnamefont
  {Shapiro}},\ }\bibfield  {title} {\enquote {\bibinfo {title} {Distributed
  quantum sensing using continuous-variable multipartite entanglement},}\
  }\href {\doibase 10.1103/PhysRevA.97.032329} {\bibfield  {journal} {\bibinfo
  {journal} {Phys. Rev. A}\ }\textbf {\bibinfo {volume} {97}},\ \bibinfo
  {pages} {032329} (\bibinfo {year} {2018})}\BibitemShut {NoStop}%
\bibitem [{\citenamefont {Xia}\ \emph {et~al.}(2020)\citenamefont {Xia},
  \citenamefont {Li}, \citenamefont {Clark}, \citenamefont {Hart},
  \citenamefont {Zhuang},\ and\ \citenamefont {Zhang}}]{Xia2020}%
  \BibitemOpen
  \bibfield  {author} {\bibinfo {author} {\bibfnamefont {Yi}~\bibnamefont
  {Xia}}, \bibinfo {author} {\bibfnamefont {Wei}\ \bibnamefont {Li}}, \bibinfo
  {author} {\bibfnamefont {William}\ \bibnamefont {Clark}}, \bibinfo {author}
  {\bibfnamefont {Darlene}\ \bibnamefont {Hart}}, \bibinfo {author}
  {\bibfnamefont {Quntao}\ \bibnamefont {Zhuang}}, \ and\ \bibinfo {author}
  {\bibfnamefont {Zheshen}\ \bibnamefont {Zhang}},\ }\bibfield  {title}
  {\enquote {\bibinfo {title} {Demonstration of a reconfigurable entangled
  radio-frequency photonic sensor network},}\ }\href {\doibase
  10.1103/PhysRevLett.124.150502} {\bibfield  {journal} {\bibinfo  {journal}
  {Phys. Rev. Lett.}\ }\textbf {\bibinfo {volume} {124}},\ \bibinfo {pages}
  {150502} (\bibinfo {year} {2020})}\BibitemShut {NoStop}%
\bibitem [{\citenamefont {Guo}\ \emph {et~al.}(2020)\citenamefont {Guo},
  \citenamefont {Breum}, \citenamefont {Borregaard}, \citenamefont {Izumi},
  \citenamefont {Larsen}, \citenamefont {Gehring}, \citenamefont {Christandl},
  \citenamefont {Neergaard-Nielsen},\ and\ \citenamefont {Andersen}}]{Guo2020}%
  \BibitemOpen
  \bibfield  {author} {\bibinfo {author} {\bibfnamefont {Xueshi}\ \bibnamefont
  {Guo}}, \bibinfo {author} {\bibfnamefont {Casper~R.}\ \bibnamefont {Breum}},
  \bibinfo {author} {\bibfnamefont {Johannes}\ \bibnamefont {Borregaard}},
  \bibinfo {author} {\bibfnamefont {Shuro}\ \bibnamefont {Izumi}}, \bibinfo
  {author} {\bibfnamefont {Mikkel~V.}\ \bibnamefont {Larsen}}, \bibinfo
  {author} {\bibfnamefont {Tobias}\ \bibnamefont {Gehring}}, \bibinfo {author}
  {\bibfnamefont {Matthias}\ \bibnamefont {Christandl}}, \bibinfo {author}
  {\bibfnamefont {Jonas~S.}\ \bibnamefont {Neergaard-Nielsen}}, \ and\ \bibinfo
  {author} {\bibfnamefont {Ulrik~L.}\ \bibnamefont {Andersen}},\ }\bibfield
  {title} {\enquote {\bibinfo {title} {Distributed quantum sensing in a
  continuous-variable entangled network},}\ }\href {\doibase
  10.1038/s41567-019-0743-x} {\bibfield  {journal} {\bibinfo  {journal} {Nature
  Physics}\ }\textbf {\bibinfo {volume} {16}},\ \bibinfo {pages} {281--284}
  (\bibinfo {year} {2020})}\BibitemShut {NoStop}%
\bibitem [{\citenamefont {Cram{\'e}r}(1999)}]{cramer1999mathematical}%
  \BibitemOpen
  \bibfield  {author} {\bibinfo {author} {\bibfnamefont {Harald}\ \bibnamefont
  {Cram{\'e}r}},\ }\href@noop {} {\emph {\bibinfo {title} {Mathematical methods
  of statistics}}},\ Vol.~\bibinfo {volume} {9}\ (\bibinfo  {publisher}
  {Princeton university press},\ \bibinfo {year} {1999})\BibitemShut {NoStop}%
\bibitem [{\citenamefont {Olivares}\ and\ \citenamefont
  {Paris}(2009)}]{olivares2009}%
  \BibitemOpen
  \bibfield  {author} {\bibinfo {author} {\bibfnamefont {Stefano}\ \bibnamefont
  {Olivares}}\ and\ \bibinfo {author} {\bibfnamefont {Matteo~GA}\ \bibnamefont
  {Paris}},\ }\bibfield  {title} {\enquote {\bibinfo {title} {Bayesian
  estimation in homodyne interferometry},}\ }\href@noop {} {\bibfield
  {journal} {\bibinfo  {journal} {Journal of Physics B: Atomic, Molecular and
  Optical Physics}\ }\textbf {\bibinfo {volume} {42}},\ \bibinfo {pages}
  {055506} (\bibinfo {year} {2009})}\BibitemShut {NoStop}%
\bibitem [{\citenamefont {Berni}\ \emph {et~al.}(2015)\citenamefont {Berni},
  \citenamefont {Gehring}, \citenamefont {Nielsen}, \citenamefont
  {H{\"a}ndchen}, \citenamefont {Paris},\ and\ \citenamefont
  {Andersen}}]{berni2015}%
  \BibitemOpen
  \bibfield  {author} {\bibinfo {author} {\bibfnamefont {Adriano~A}\
  \bibnamefont {Berni}}, \bibinfo {author} {\bibfnamefont {Tobias}\
  \bibnamefont {Gehring}}, \bibinfo {author} {\bibfnamefont {Bo~M}\
  \bibnamefont {Nielsen}}, \bibinfo {author} {\bibfnamefont {Vitus}\
  \bibnamefont {H{\"a}ndchen}}, \bibinfo {author} {\bibfnamefont {Matteo~GA}\
  \bibnamefont {Paris}}, \ and\ \bibinfo {author} {\bibfnamefont {Ulrik~L}\
  \bibnamefont {Andersen}},\ }\bibfield  {title} {\enquote {\bibinfo {title}
  {Ab initio quantum-enhanced optical phase estimation using real-time feedback
  control},}\ }\href@noop {} {\bibfield  {journal} {\bibinfo  {journal} {Nature
  Photonics}\ }\textbf {\bibinfo {volume} {9}},\ \bibinfo {pages} {577--581}
  (\bibinfo {year} {2015})}\BibitemShut {NoStop}%
\bibitem [{\citenamefont {{Stoica}}\ and\ \citenamefont
  {{Marzetta}}(2001)}]{Stoica2001}%
  \BibitemOpen
  \bibfield  {author} {\bibinfo {author} {\bibfnamefont {P.}~\bibnamefont
  {{Stoica}}}\ and\ \bibinfo {author} {\bibfnamefont {T.~L.}\ \bibnamefont
  {{Marzetta}}},\ }\bibfield  {title} {\enquote {\bibinfo {title} {Parameter
  estimation problems with singular information matrices},}\ }\href@noop {}
  {\bibfield  {journal} {\bibinfo  {journal} {IEEE Transactions on Signal
  Processing}\ }\textbf {\bibinfo {volume} {49}},\ \bibinfo {pages} {87--90}
  (\bibinfo {year} {2001})}\BibitemShut {NoStop}%
\bibitem [{\citenamefont {Gross}\ and\ \citenamefont
  {Caves}(2020)}]{Gross2020}%
  \BibitemOpen
  \bibfield  {author} {\bibinfo {author} {\bibfnamefont {Jonathan~Arthur}\
  \bibnamefont {Gross}}\ and\ \bibinfo {author} {\bibfnamefont {Carlton~M}\
  \bibnamefont {Caves}},\ }\bibfield  {title} {\enquote {\bibinfo {title} {One
  from many: Estimating a function of many parameters},}\ }\href {\doibase
  10.1088/1751-8121/abb9ed} {\bibfield  {journal} {\bibinfo  {journal} {Journal
  of Physics A: Mathematical and Theoretical}\ } (\bibinfo {year} {2020}),\
  10.1088/1751-8121/abb9ed}\BibitemShut {NoStop}%
\end{thebibliography}%

\end{document}